\documentclass{article}
\usepackage{spconf,amsmath,graphicx}
\usepackage{subfloat} 
\usepackage{algorithmic}
\usepackage{array}
\usepackage{eqparbox}
\usepackage{caption}
\usepackage[font=footnotesize]{subfig}
\usepackage{fixltx2e}
\usepackage{multirow}
\usepackage{url}
\usepackage{booktabs}

\title{A GMM-based Stair Quality Model for Human Perceived JPEG Images}
\name{Sudeng Hu, Haiqiang Wang and C.-C. Jay Kuo}         
\address{Media Communications Laboratory, University of Southern California}
\begin{document}
\ninept

\maketitle

\begin{abstract}
Based on the notion of just noticeable differences (JND), a stair
quality function (SQF) was recently proposed to model human perception
on JPEG images. Furthermore, a k-means clustering algorithm was adopted
to aggregate JND data collected from multiple subjects to generate a
single SQF.  In this work, we propose a new method to derive the SQF
using the Gaussian Mixture Model (GMM).  The newly derived SQF can be
interpreted as a way to characterize the mean viewer experience.
Furthermore, it has a lower information criterion (BIC) value than the
previous one, indicating that it offers a better model. A specific
example is given to demonstrate the advantages of the new approach. 
\end{abstract}
\section{Introduction}\label{sec:intro}

The traditional quality metric for coded image/video, such as the
peak signal-to-noise ratio (PSNR), is a continuous function of the coding bit
rate \cite{chandler2007vsnr,cit:HVS_IMQ3, cit:TCSVT_2006_CSF_Quality,
movie}. Several newly proposed quality metrics such as SSIM
\cite{wang2004image} and perceptually weighted PSNR \cite{hu} still
preserve this property. However, these continuous quality models
contradict to our subjective visual experience since human can only
differentiate a small number of quality levels. 

Based on the notion of just noticeable difference (JND), it was shown in
\cite{MCLJND} that human-perceived quality of JPEG images is a stair
function of the quality factor (QF).  It is a monotonically increasing
piecewise-constant function characterized by a couple of jumps.  The
stair quality function (SQF) is discontinuous, and its jumps can be
interpreted as the JND points between two adjacent quality levels.  For
a given image coded by JPEG with multiple QFs, the number of discrete
quality levels and the location of JND points vary among test subjects.
Since they are random variables, it is important to develop a
methodology to integrate the data collected from multiple test subjects.
A simple k-means clustering algorithm was proposed in \cite{MCLJND} to
process collected JND data to generate the aggregate SQF, which is
called the K-SQF here. 

In this work, we treat JND points from a subject as samples, and use the
Gaussian Mixture Model (GMM) to fit the sample distribution. This
approach leads to another aggregate SQF, called the G-SQF.  The G-SQF
can be interpreted as the mean viewer experience in differentiating
compressed image quality under a wide range of coding bit rates. The
shape parameters of the G-SQF such as the number of discrete quality
levels and the location and height of JND points are determined
automatically by this modeling procedure. As compared to the ad hoc
k-means clustering algorithm used in deriving the K-SQF, the G-SQF is
rooted in solid theoretical foundation.  We will show that the G-SQF has
a lower information criterion (BIC) value than the K-SQF, indicating
that it is a better model. Furthermore, a specific example will be given
to demonstrate the advantages of G-SQF over K-SQF. 

This rest of this paper is organized as follows. The data collection
procedure for JND-based subjective JPEG quality assessment is reviewed
in Section 2.  A GMM-based processing technique is proposed to handle
collected JND data in Section 3. The performance comparison of G-SQF and
K-SQF is conducted in Section 4. Finally, concluding remarks are given
in Section 5. 

\section{JND-based Subjective JPEG Quality Assessment}\label{sec:format}

The process of building a large-scale human-centric quality dataset for
JPEG-coded images, called MCL-JCI, is described below. MCL-JCI contains
50 source (or uncompressed) images of resolution 1920x1080. Each source
image is coded by the JPEG encoder \cite{JPEG} 100 times with the quality
factor (QF) set from 1, 2, 3 ... to 100. Thus, the whole MCL-JCI
dataset consists of 5,050 images in total. The quality of coded images
with respect to each source image is evaluated by 20 subjects. They were
seated in a controlled environment. The viewing distance was 2 meters
(1.6 times the picture height) from the center of the monitor to the
seat. The image pair was displayed on a 65" TV with native resolution of
3840x2160. A subject compared two images displayed side by side and
determined whether these two images are noticeably different. 

The following bisection search procedure was adopted to offer a more
robust and efficient pairwise comparison result. 
\begin{itemize}
\item {\bf Initialization.} We begin with comparing images of the best
and the worst quality. The best quality is obtained by setting QF=100
while the worst quality is set to the QF value that gives the lowest
acceptable through subjective test. Before the subjective test, a small
number of volunteers were asked to find the lowest acceptable QF
parameter. 
\item {\bf Iteration.} Compare two images whose QF is located at two
ends of the interval of interest and see whether they have noticeable
difference or not. If no, no further search is needed for this interval
since it does not contain a JND point. If yes, we partition the interval
into two halves of equal length, and repeat the same comparison
procedure iteratively until one of the two termination criteria is
reached. 
\item {\bf Termination.} There are two termination cases. First, the
process is terminated when the interval length reaches the minimum value
with the QF difference equal to one. Second, one observes noticeable
difference at a certain level and cannot observe any noticeable
difference at the next level. 
\end{itemize}

\begin{figure}[tb]
\centering
\subfloat[]{\includegraphics[width=0.25\textwidth]{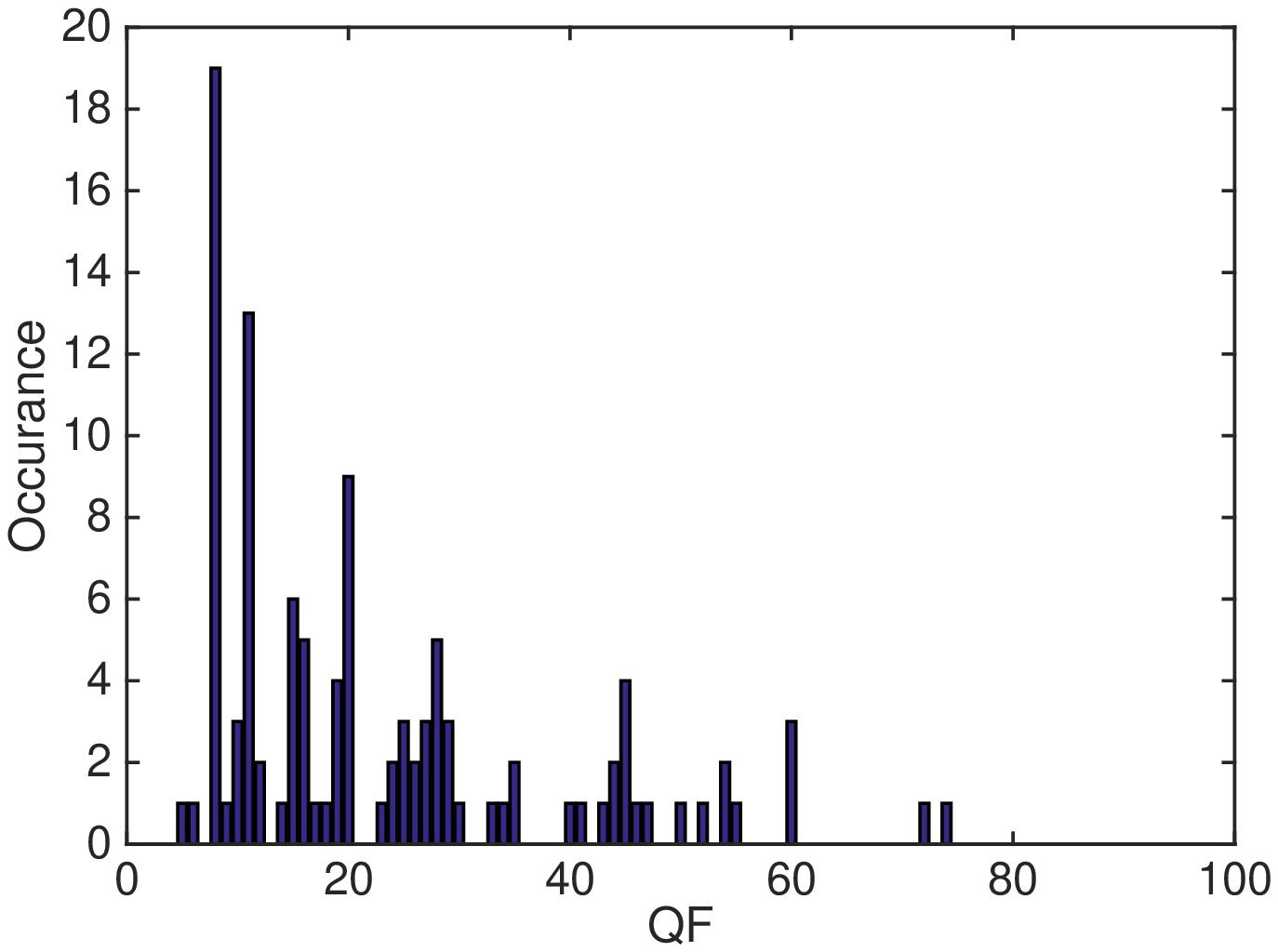}}
\subfloat[]{\includegraphics[width=0.25\textwidth]{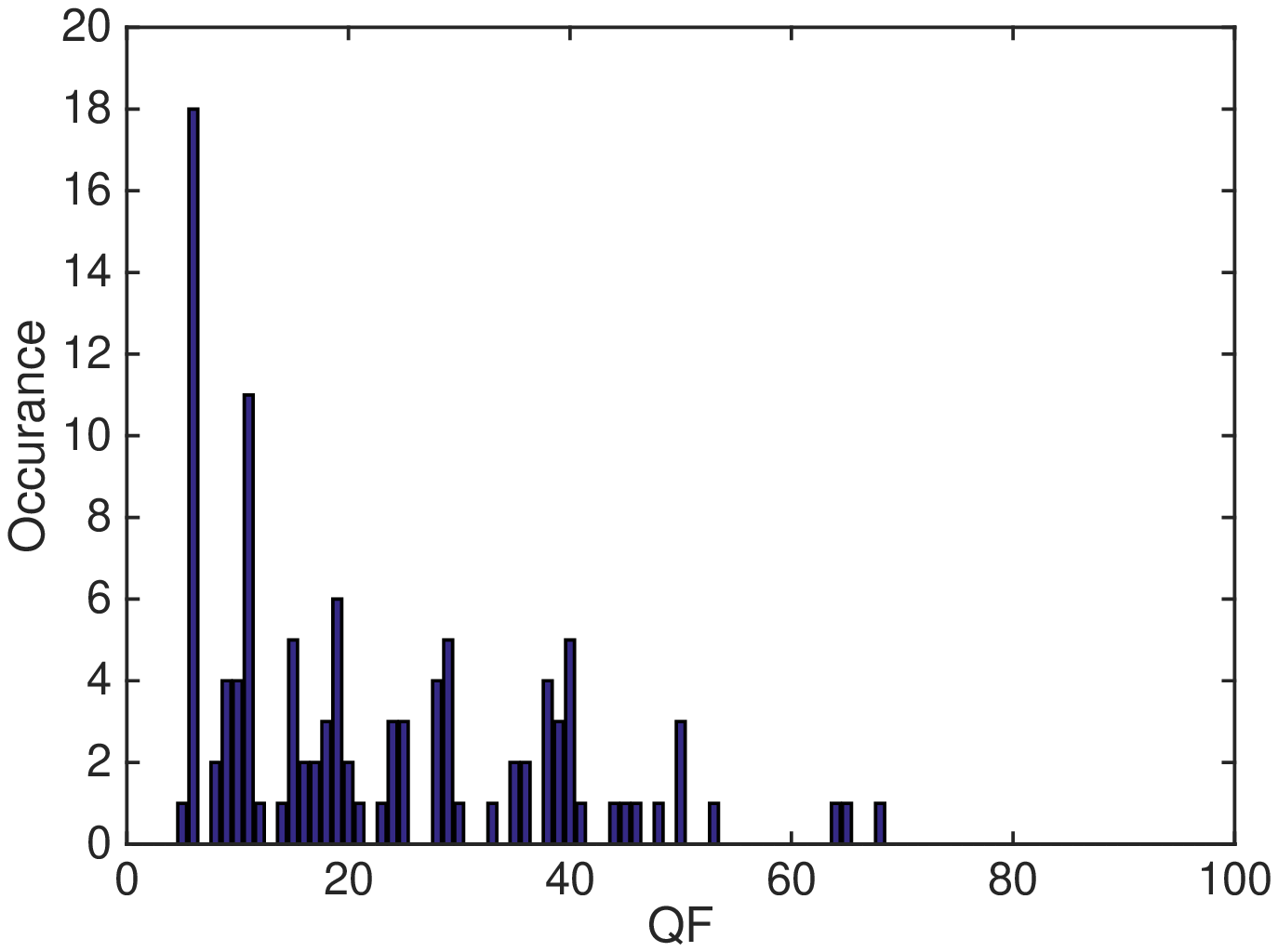}} 
\caption{The histograms of JND points for source image (a) No. 06 
and (b) No. 26.}\label{fig:histogram}
\end{figure}

The above subjective test produces raw JND data samples for each image,
where one subject contributes a set of JND samples. The histograms of
JND points for two exemplary images are shown in Fig.
\ref{fig:histogram}, where (a) and (b) are obtained from source image
No. 6 and No. 26, respectively. They are too complicated to be used as
is. It is essential to process them and build an aggregate SQF for each
individual image. Ideally, the aggregate SQF can be used to characterize
the mean experience of subjects in the test. The derivation of an
accurate SQF will facilitate the use of the machine learning technique
in predicting the SQF for images not in the dataset. 

The k-means clustering algorithm was proposed in \cite{MCLJND} to
process collected JND samples for an aggregate SQF, where $k$ is
determined by the rounded mean of distinguishable quality levels.
However, the k-means clustering approach is a heuristic one. It is
difficult to give the resulting SQF any statistical meaning.  To address
this shortcoming, we propose a new method to process collected JND
samples based on the GMM. Mathematically, it is easy to see that the
resulting SQF offers the mean viewer experience among all subjects
participating in the test. 

\section{GMM-based JND Data Processing}

\subsection{Group Partitioning of JND Points}

Images coded in the range of high QFs have good perceptual quality and
their distortion can hardly be perceived. As a result, there are only
few JND points falling in this range as compare to those in the low QF
range.  This phenomenon is obvious in the exemplary histograms given in
Fig. \ref{fig:histogram}.  Because perceptual difference in high quality
images is so small that it can be easily neglected when compared with
low quality images in statistical analysis.  Furthermore, compressed
images with high QF values are more important in practice since people
are interested in high quality images in most applications.  They are
much more frequently used than those compressed with low QF values. For
the above-mentioned reasons, the high QF JND points should not be merged
with the low QF JND points to form components in one single GMM.  By
following a similar argument, one can argue that JND points in the low
QF range should not be merged with those in the middle QF range.  As a
result, we classify JND samples into three main groups according to
their locations: high QF, middle QF and low QF groups. This is achieved
by a partitioning scheme described below. 

\begin{figure}[tb]
\centering
\includegraphics[width=0.3\textwidth]{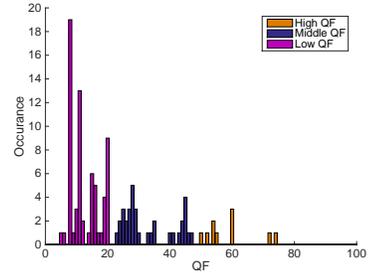} 
\caption{Partitioning of JND points into high, middle and low QF
groups, where each group will be modeled by a GMM independently.} 
\label{fig:combinedHisto}
\end{figure}

First, we order JND points according to their QF values (from the
largest to the smallest) and identify the JND points lying at the top
10\% and 50\% locations.  Then, we examine the heights of these two JND
points against the JND histogram curve.  There are three scenarios. 
\begin{enumerate}
\item If a JND point happens to be a local minimum (or zero), we select
it as the boundary point between two groups. 
\item If a JND point is a local maximum, we search two local minima (or
zero) along its left and right directions and select the smaller one as
the updated boundary point. 
\item If the JND point is neither a local maximum nor a local minimum,
we search along the descending direction for the local minimum (or zero)
and select it as the updated boundary point. 
\end{enumerate}
We split the height of a boundary point (i.e. its number count) equally
into two halves - one goes to the left and the other goes to the right.
By following the above steps, we obtain the high, middle and low QF
groups. An example is given in Fig. \ref{fig:combinedHisto}.  Then, we
will use three GMMs to model their JND distributions independently. 

\subsection{JND Histogram Modeling with GMM}

The number of distinguishable quality levels and their JND positions
depend on both image content and test subjects.  Even for the same image
and the same QF group, it is still difficult to group JND points since
different subjects may have different numbers of JND points.  To proceed
with statistical analysis, we need an underlying model for the JND
distribution. Here, it is assumed that the JND distribution is in form
of GMM with $N$ components.  Mathematically, it can be expressed as
\begin{equation}\label{eq:GMM}
f(x) =  \sum_{i=1}^N  \alpha_i \cdot \frac{1}{\sqrt{2\pi}\sigma_i}
exp({-\frac{(x-q_i)^2}{2\sigma_i^2}}), 
\end{equation}
where each component is a normal distribution with mean $q_i$ and
variance $\sigma_i^2$, and $a_i$ is the mixture weight satisfying the
constraint $\sum_{i=1}^N\alpha_i=1$. 

To determine the set of parameters of GMM in Eq.  (\ref{eq:GMM}); namely,
$$
\Theta=\{\alpha_i, q_i, \sigma_i\}, \quad i=1,\cdots, N,
$$ 
we adopt the Expectation Maximization (EM) algorithm
\cite{dempster1977maximum}.  It is well known that the EM algorithm is
an iterative algorithm that updates these parameters in each iteration
until the process converges or reaches the preset maximum iteration
number. The EM algorithm is sensitive to the initial values of these
parameters.  In the proposed method, we compute the histogram of JND
samples in the target QF region, and select the location of $N$ largest
bins as initial value for the mean of $N$ components (i.e., $q_i$,
$i=1,\cdots, N$).  The initial variance of all components is set to 1. 
 
Furthermore, we need to specify the component number, $N$, of the GMM.
If $N$ is too small, it is difficult to fit the JND samples well. If $N$
is too large, it may result in overfit.  Here, we perform an exhaustive
search for the optimal component number $N^*$. That is, we begin with
$N=1$, and increase its value by one every time until $N$ reaches the
pre-set maximum component number of each group.  We have the following
observation based on a large number of experiments.  For the high QF
group, the optimal $N^*$ is either one or two so that the maximum $N$ is
set to three. Similarly, the maximum component numbers are set to four
and three for the middle and the low QF groups, respectively. Thus, the
cost of exhaustive search is under control. 

We use the Bayesian information criterion (BIC)
\cite{findley1991counterexamples} to determine the best GMM. A lower BIC
value indicates better performance.  Mathematically, the BIC is defined as
\begin{equation}\label{eq:BIC}
\mbox{BIC} = -2 \cdot ln(\hat{L}) + k \cdot ln(n),
\end{equation}
where $\hat{L}$ is the maximized value of the likelihood function of the
model, $ln$ is the natural log, $k$ is the number of free parameters in
the model, and $n$ is the number of samples. In the current case,
$$
\hat{L}=p( x | \Theta ),
$$ 
where $x$ denotes all samples and $\Theta$ is the set of GMM parameters.
Both terms of BIC in Eq. (\ref{eq:BIC}) are positive. A better fit will
drive the first term lower and a smaller $k$ will drive the second term
lower for fixed $n$. Thus, the BIC value helps strike a balance between
data fitting performance and model complexity. 

\begin{figure}[t]
\centering
\subfloat[]{\includegraphics[width=0.25\textwidth]{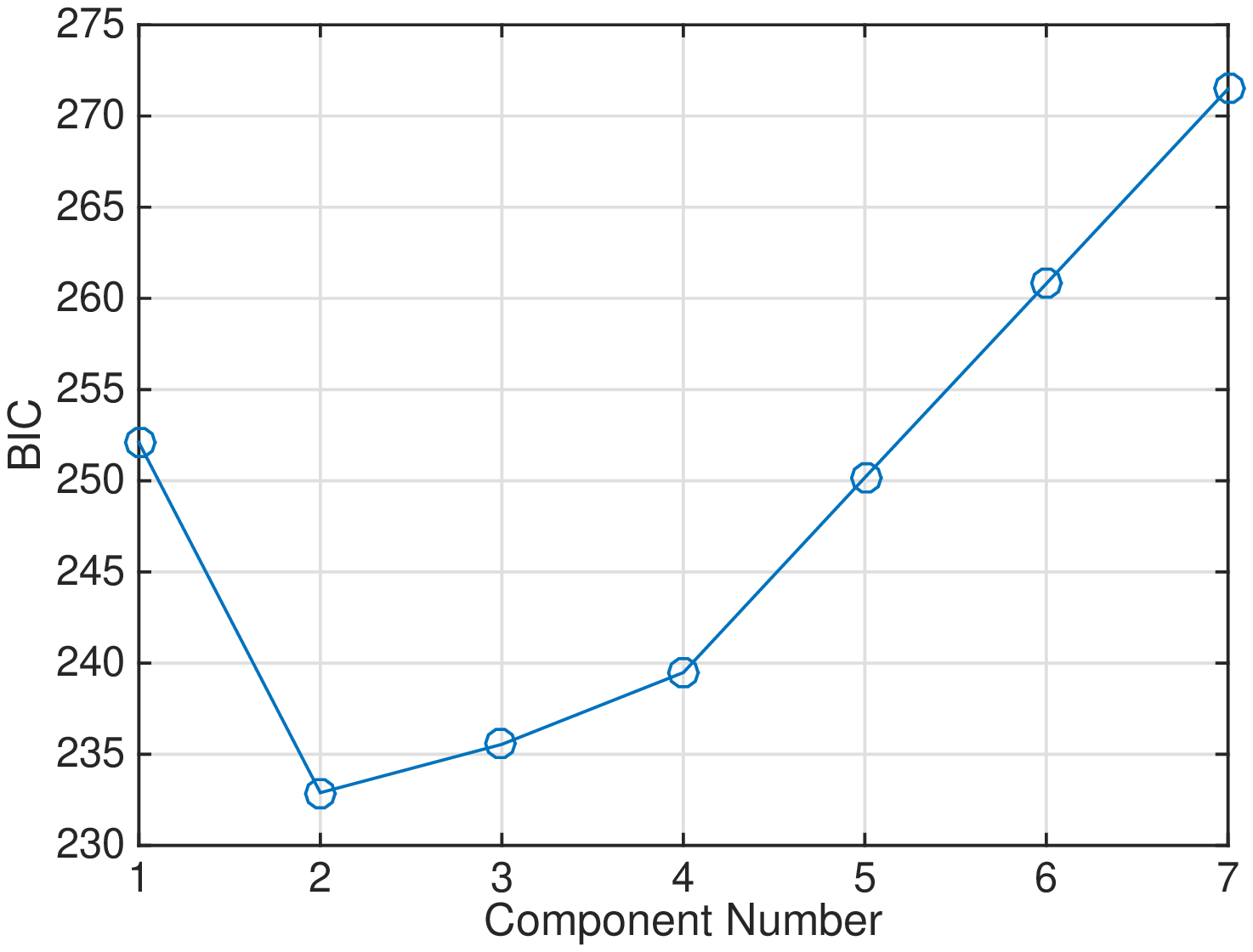}}
\subfloat[]{\includegraphics[width=0.25\textwidth]{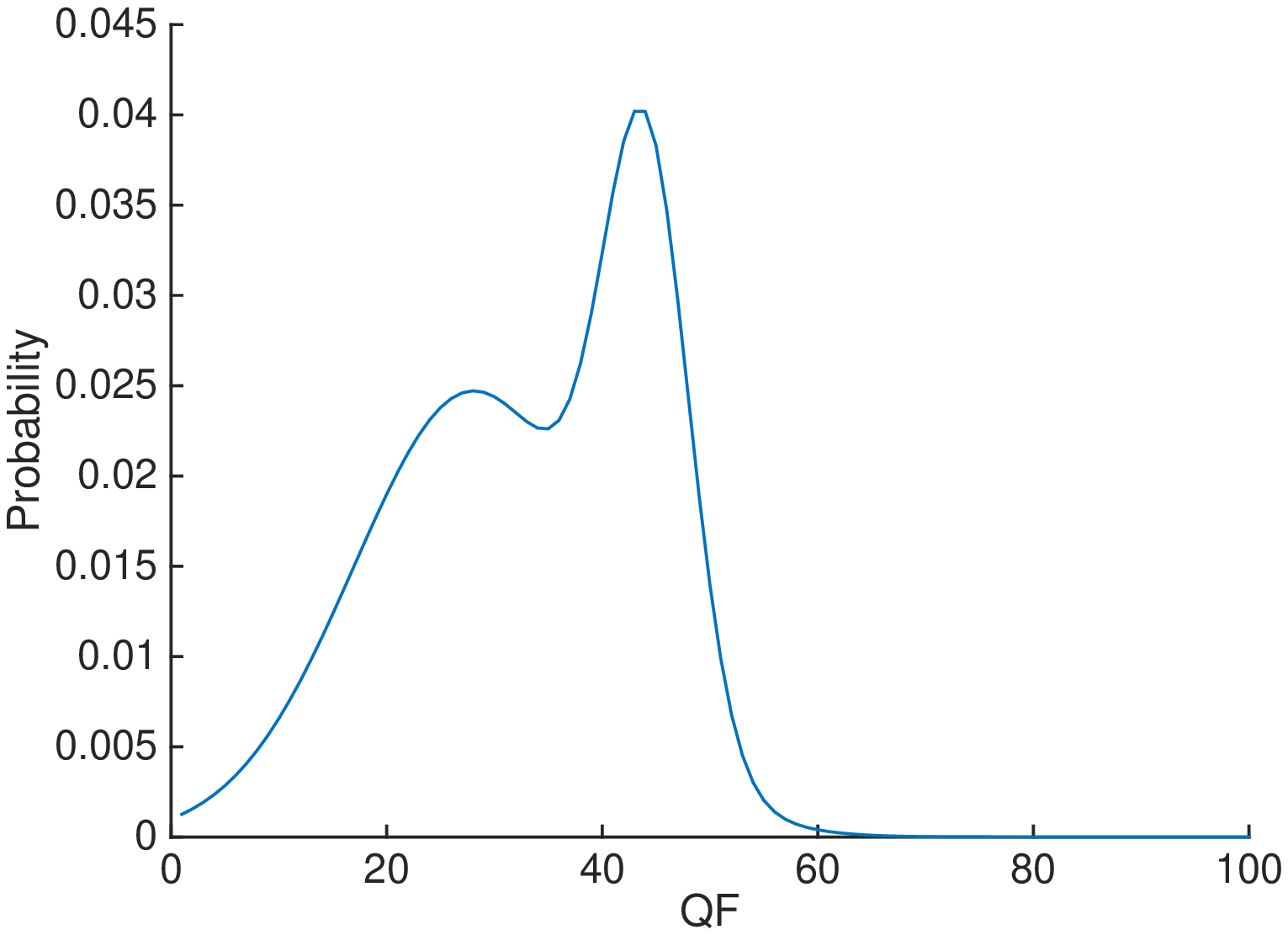}} 
\caption{Selection of the GMM component number: (a) the BIC value as a
function of the GMM component number and (b) the posterior probability 
of the optimal GMM.} \label{fig:BIC_CURVES}
\end{figure}

To give an example, for the middle QF group in Fig.
\ref{fig:combinedHisto}, we show the BIC value as a function of the GMM
component number, $N=1, \cdots, 7$, in Fig. \ref{fig:BIC_CURVES} (a).
The BIC decreases as the component number increases from $N=1$ to 2.
The BIC reaches the minimum value at $N=2$. Afterwards, the BIC
increases as $N$ increases. The probability density function of the
optimal GMM with $N=2$ is shown in Fig.  \ref{fig:BIC_CURVES} (b), where
we see two Gaussian components clearly. 

\begin{figure}[thb]
\centering
\subfloat[]{\includegraphics[width=0.25\textwidth]{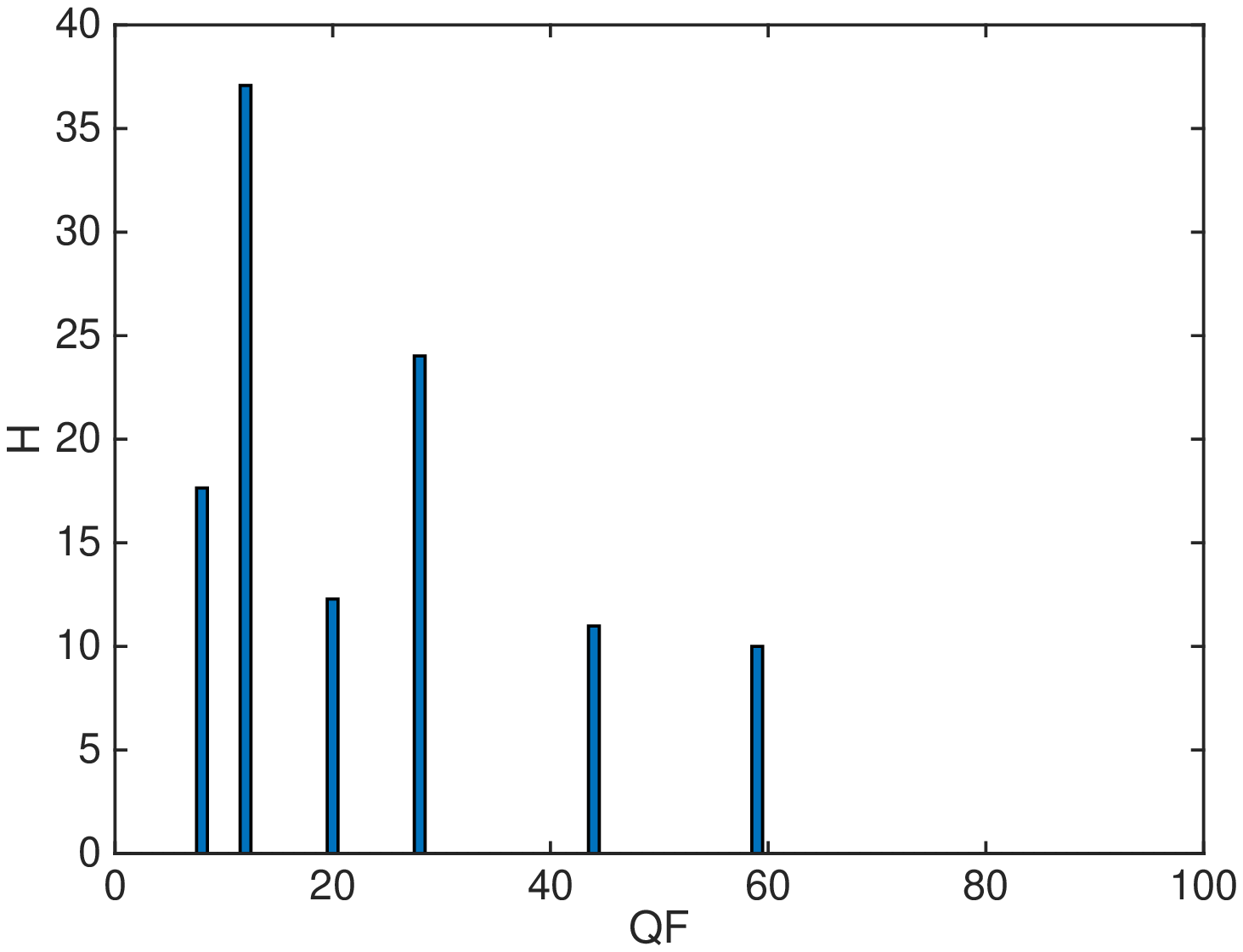}}
\subfloat[]{\includegraphics[width=0.25\textwidth]{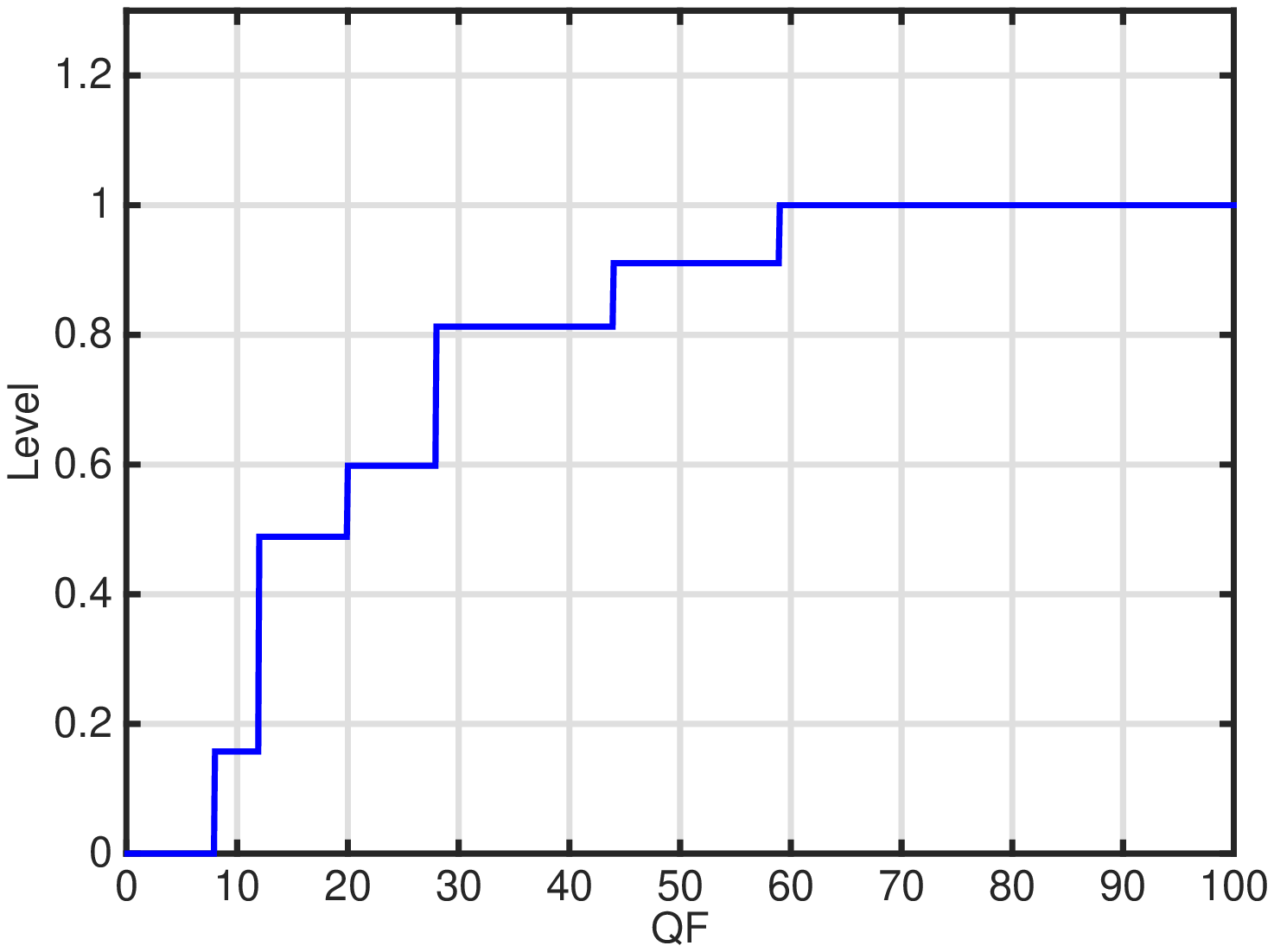}}\\
\subfloat[]{\includegraphics[width=0.25\textwidth]{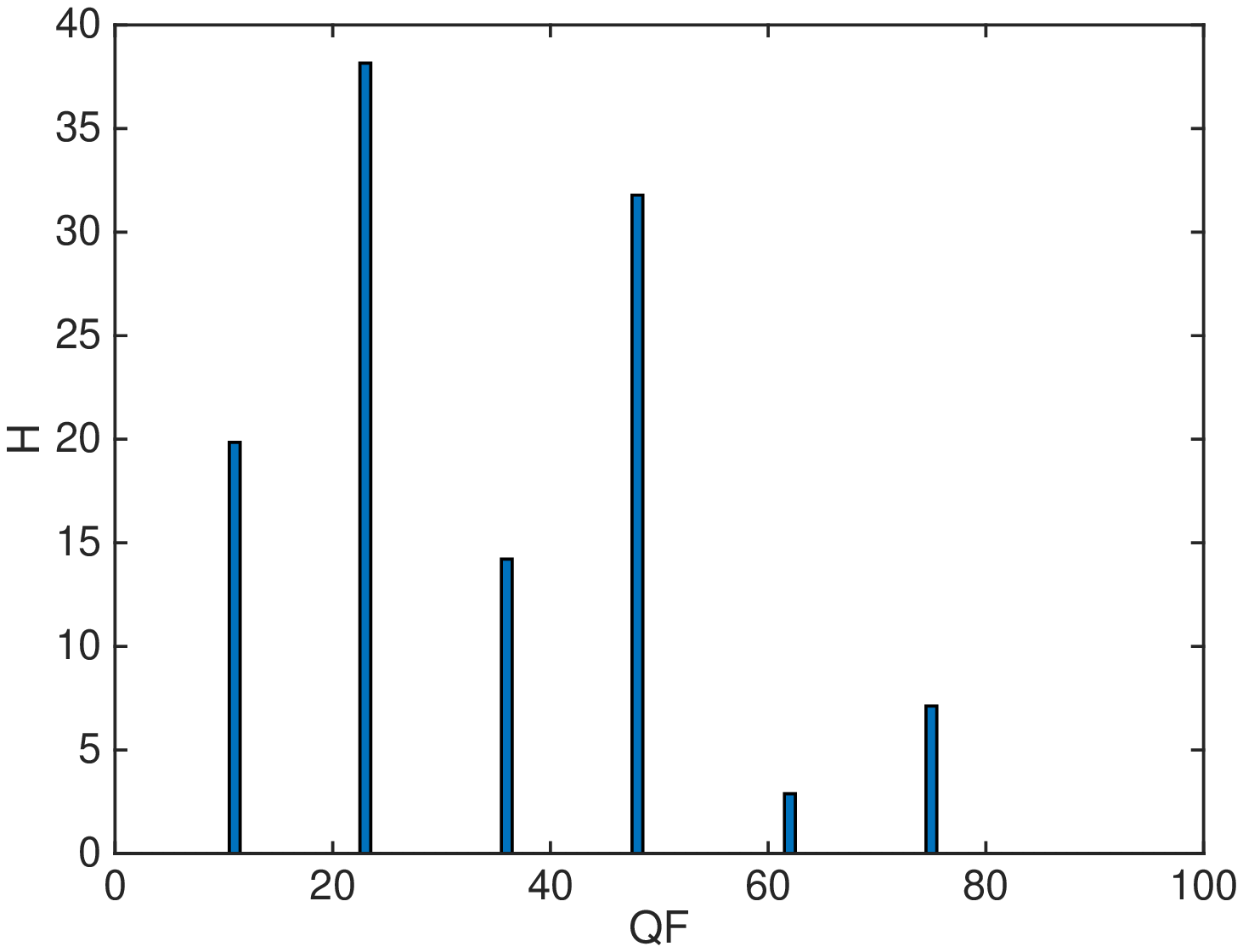}}
\subfloat[]{\includegraphics[width=0.25\textwidth]{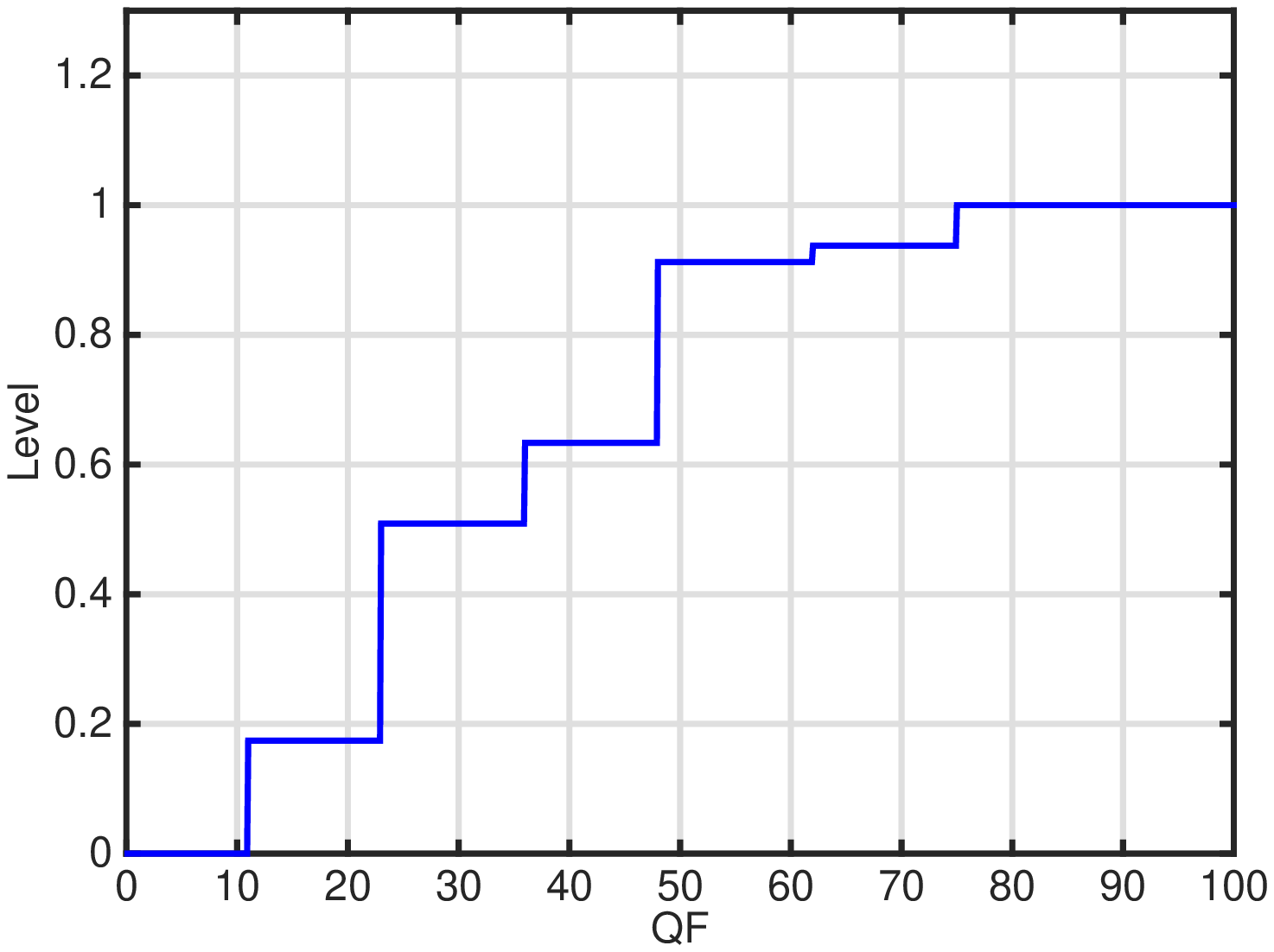}}
\caption{(a) The simplifed JND histogram of source image No. 06, (b) the
SQF of source image No. 06, (c) the simplifed JND histogram of source
image No. 26, (d) The SQF of source image No. 26.}\label{fig:jnd_cdd}
\end{figure}

\subsection{From GMM to SQF}

Once a GMM is built for each QF group (or region), the remaining task is
to build the corresponding SQF. We first discuss the SQF for a single QF
group.  Let the normal function $N(x|q_i,\sigma_i)$ be the $i$th
Gaussian function in the GMM in the corresponding region. We associate
the location of the $i$th jump in the SQF with $q_i$ while its height is
set to be proportional to the area under the weighted normal function
$\alpha_i N(x|q_i,\sigma_i)$. 

Next, we examine the SQF for all three QF regions combined together.
The JND for the whole range can be obtained by combining the three JND
sets. Mathematically, the JND function can be written as
\begin{equation}
JND(x) = \sum_{j=1}^3 \sum_{i=1}^N H_{ij} \delta(x-q_{ij}),
\end{equation}
where $\delta (\cdot)$ is the Dirac delta function and $H_{ij}$
is the percepture quality change degree at the $i$th JND position
in the $j$th group (i.e. low, middle and high QF groups).  The SQF is
the normalized cumulative sum of JND function.  Mathematically, it can
be expressed as
\begin{equation}
SQF(x) = \frac{1}{\sum_{j=1}^3 \sum_{i=1}^N H_{ij}}\int_{0}^{x} JND(t) d(t),
\end{equation}
which is a monotonically increasing piecewise constant stair function.

To give an example, after the processing of the two raw JND histograms
shown in Figs. \ref{fig:histogram} (a) and (b) obtained from 20
subjects, we can aggregate them into two simplified JND histograms as
shown in Fig.  \ref{fig:jnd_cdd} (a) and (c) while their corresponding
SQFs are shown in Figs.  \ref{fig:jnd_cdd} (b) and (d), respectively.
These two SQFs offer the mean viewer experience towards these two
images. 

\section{Experimental Results}\label{sec:result}

\begin{figure}[h]
\centering
\includegraphics[width=0.36\textwidth]{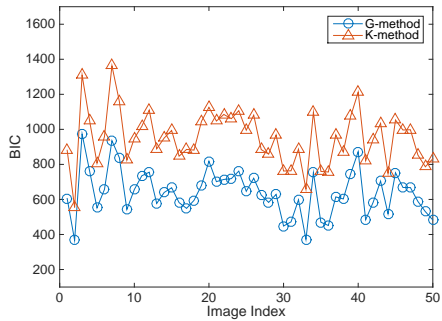} 
\caption{Comparison of the BIC values of two models with respect to 50 source images,
where the K-method and the G-method denote the k-means clustering method used in 
\cite{MCLJND} and the proposed GMM-based method, respectively.}\label{fig:BIC}
\end{figure}


\begin{figure}[tb]
\centering
\subfloat[]{\includegraphics[width=0.25\textwidth]{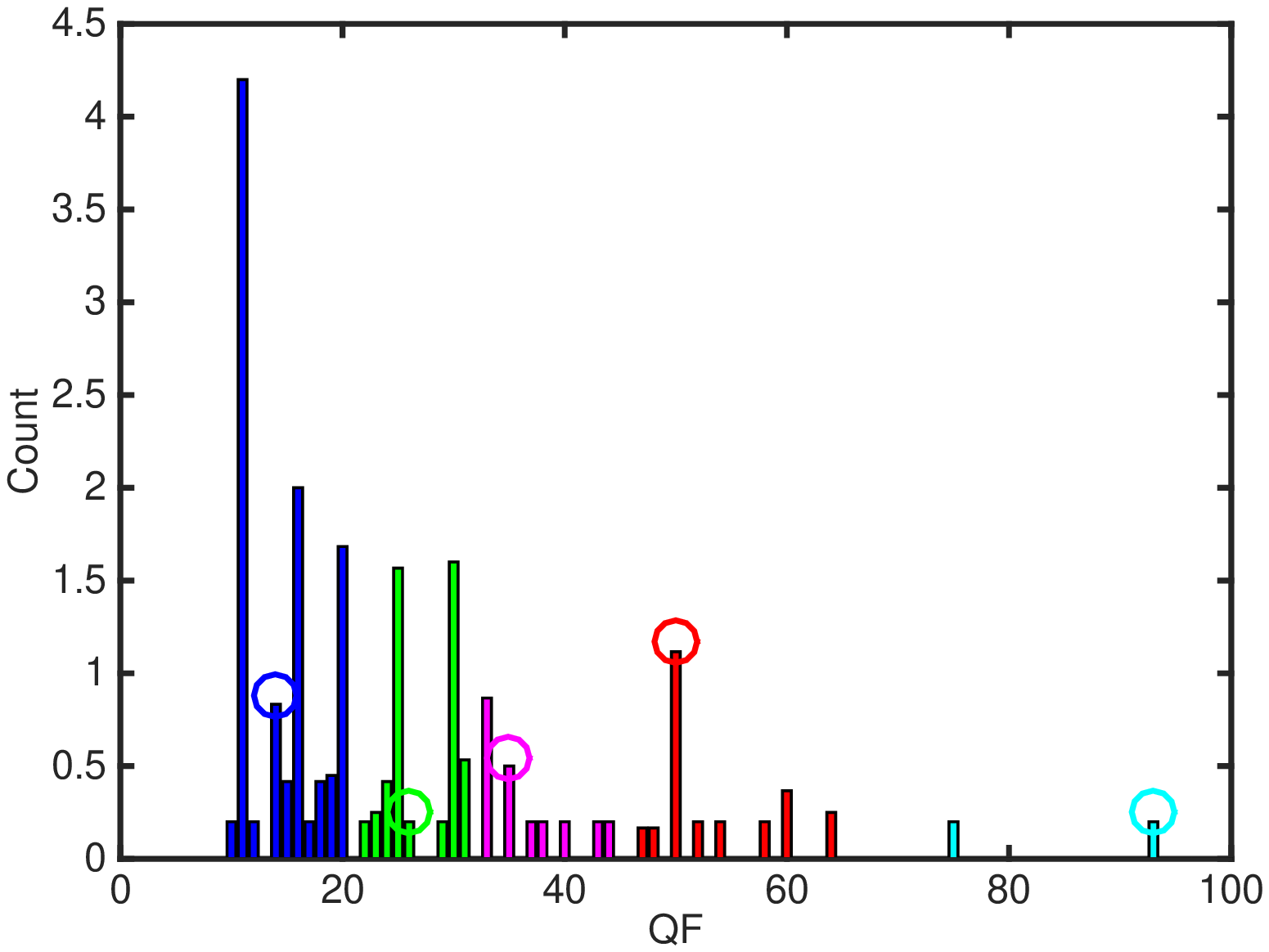}}
\subfloat[]{\includegraphics[width=0.25\textwidth]{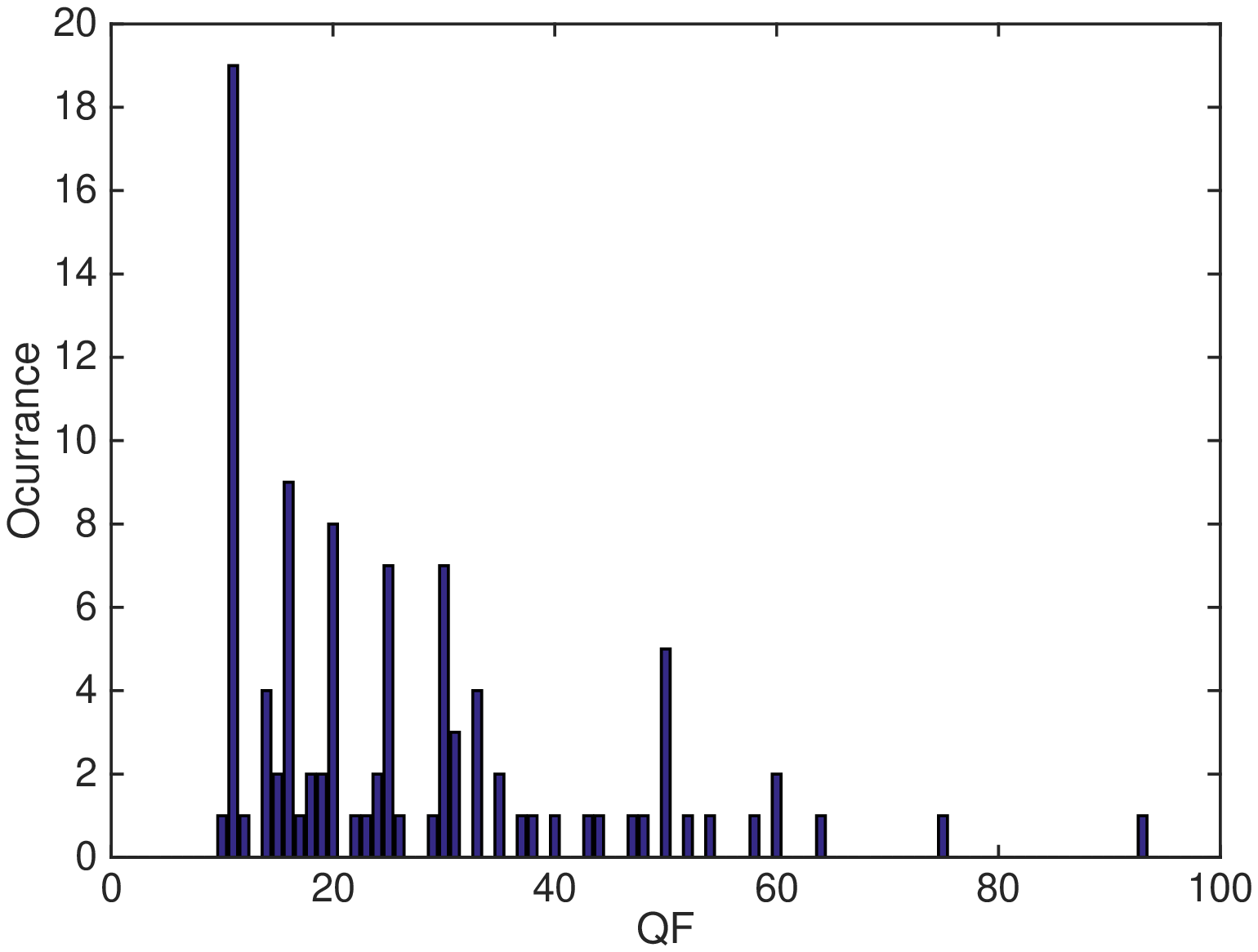}}\\
\subfloat[]{\includegraphics[width=0.25\textwidth]{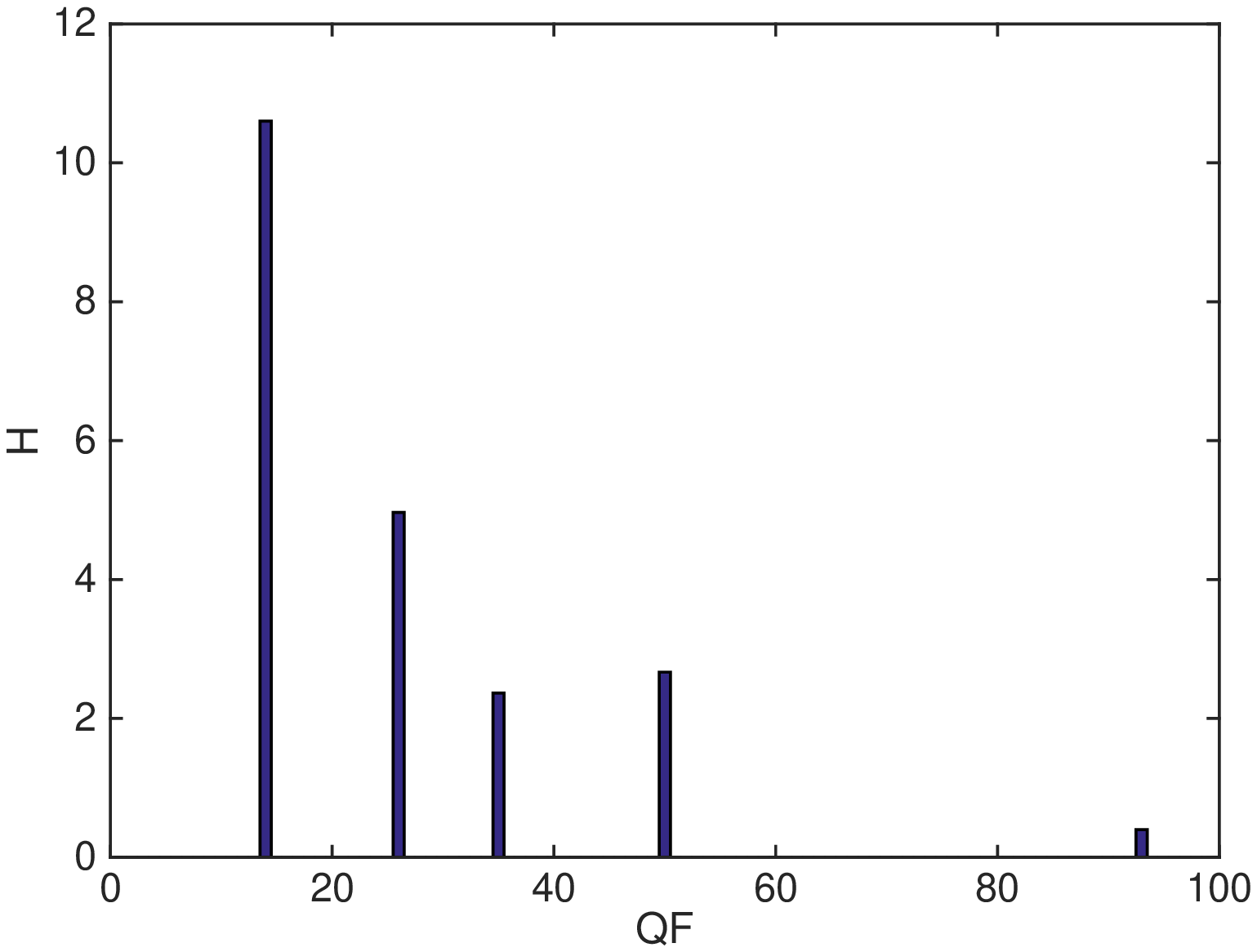}}
\subfloat[]{\includegraphics[width=0.25\textwidth]{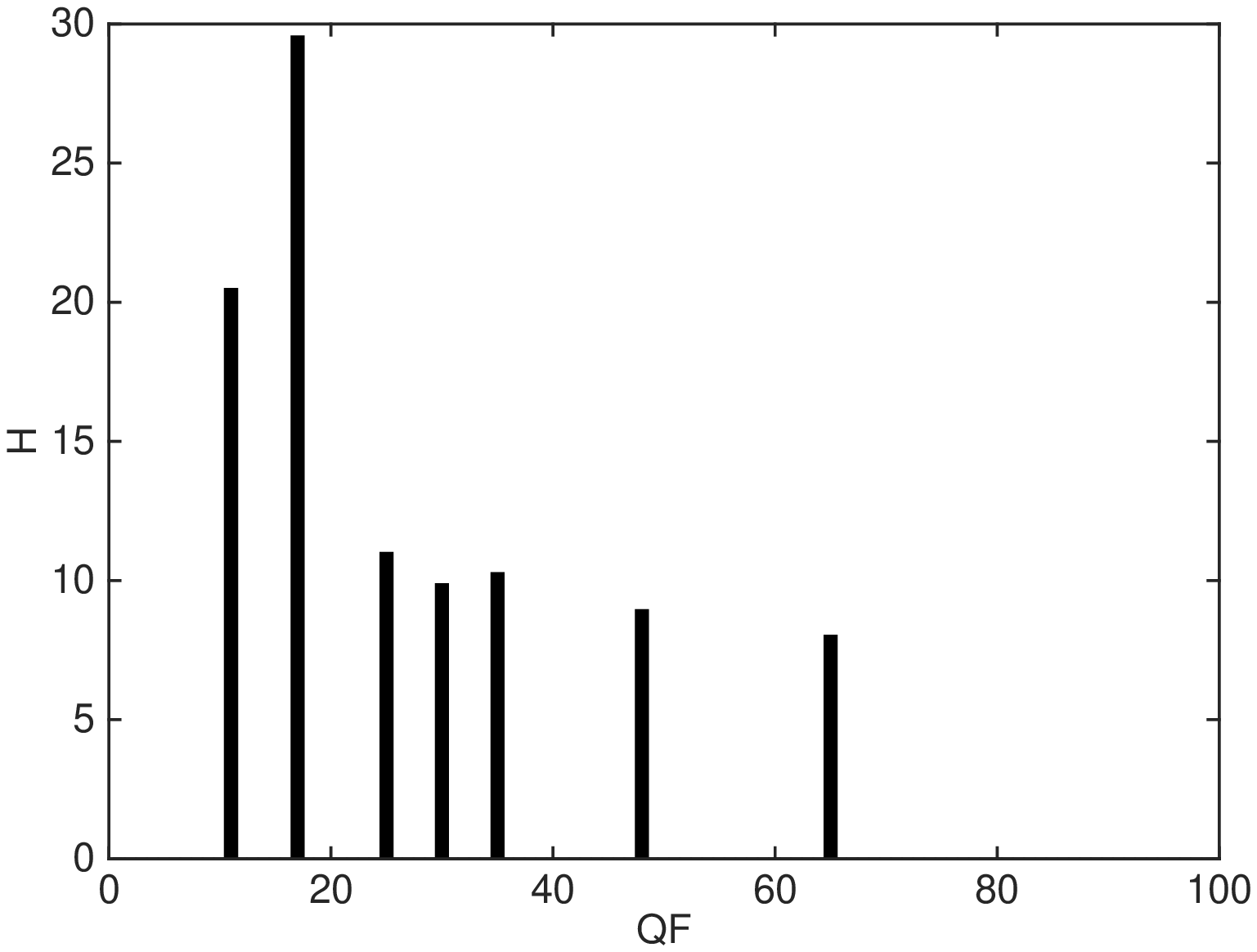}}\\
\subfloat[]{\includegraphics[width=0.25\textwidth]{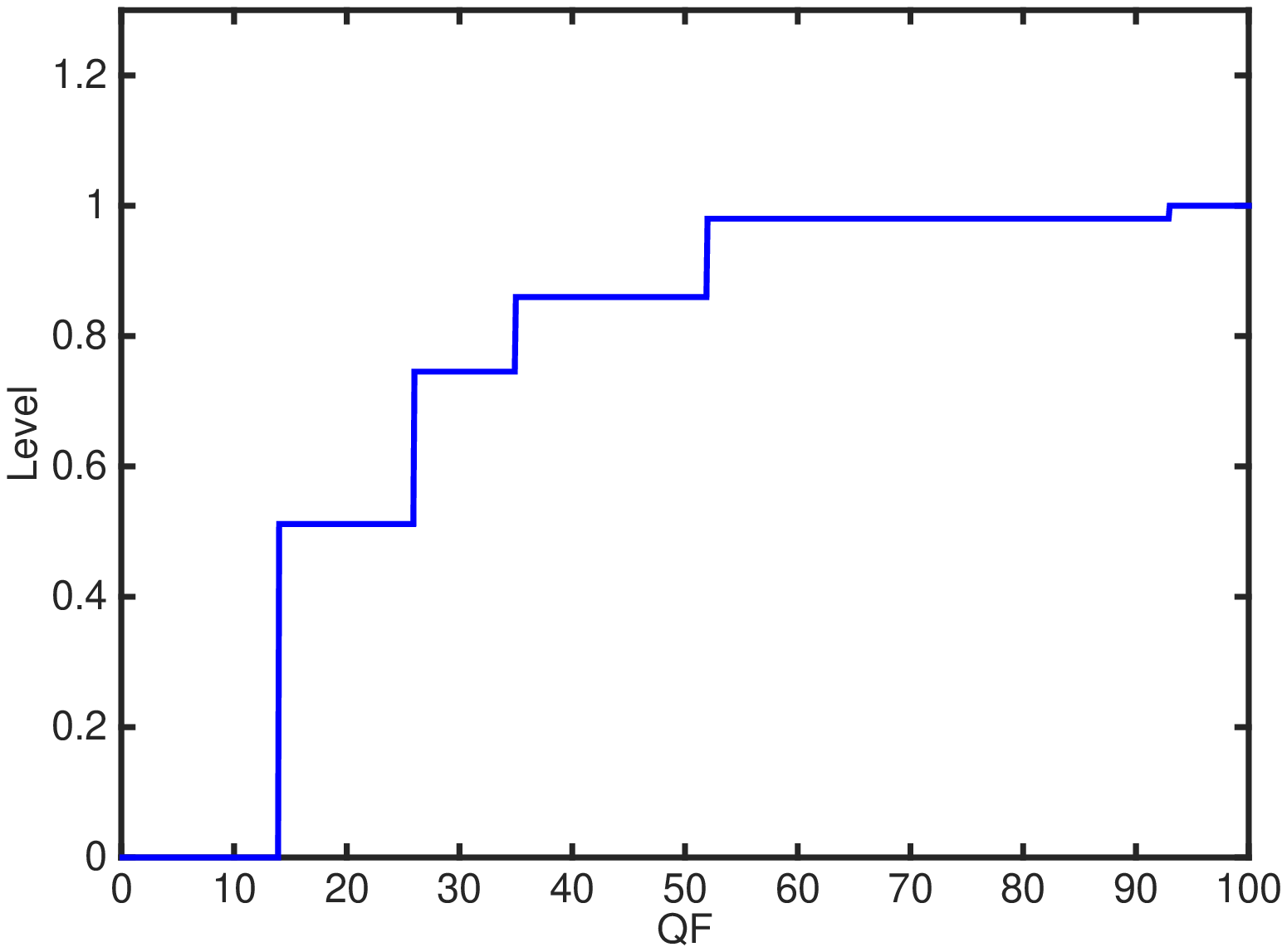}}
\subfloat[]{\includegraphics[width=0.25\textwidth]{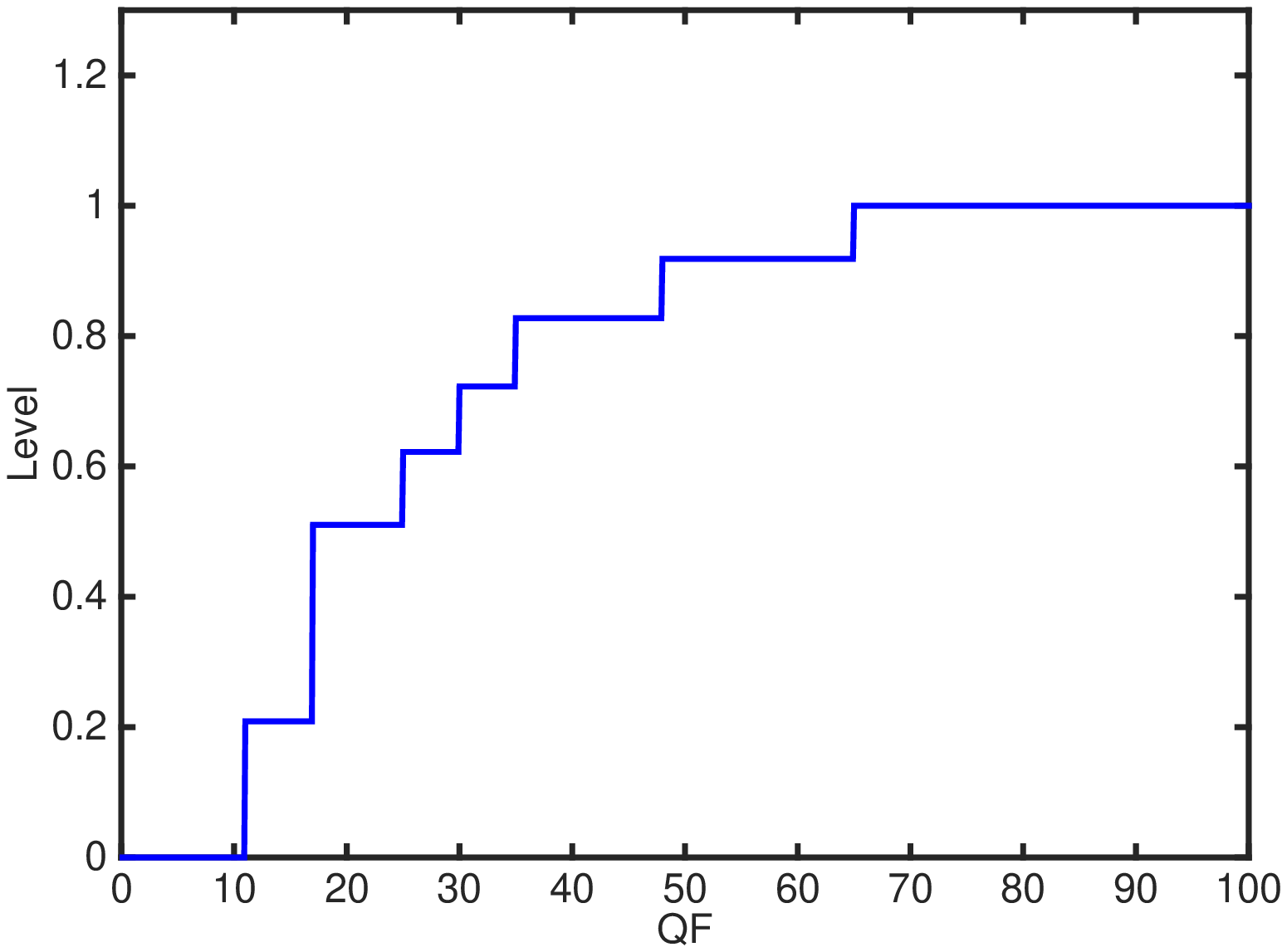}}\\
\caption{(a) The input weighted JND histogram used in the K-method, (b)
the input raw JND histogram used in the G-method, (c) the output JND
histogram processed by the K-method, (d) the output JND histogram
processed by the G-method, (e) the K-SQF, and (f) the G-SQF. }
\label{fig:jnd_example}
\end{figure}

In this section, we compare the performance of the method proposed in
\cite{MCLJND}, called the K-method, and the proposed JND data processing
method, called the G-method. The BIC in Eq.  (\ref{eq:BIC}) offers a
relative estimate of information loss in using a statistical model to
model sample data. It is used to compare the performance of two models
quantitatively. To calculate BIC of the K-method, a statistical model is
needed to model JND points.  We use a Gaussian model for each JND in the
K-method. The JND location from the K-method is set to the mean of the
Gaussian model and the model variance is set to the sample variance of
that cluster.  The BIC is calculated for all 50 images in the data set,
and the results are shown in Fig.  \ref{fig:BIC}. We see the BIC of
G-method is always lower than that of the K-method, which means the
G-method offers a better model. 

Without loss of generality, we choose source image No. 13 in Fig.
\ref{fig:BIC} as an example to shed light on the BIC values of the two
methods. The BIC value consists of two terms as presented in Eq.
(\ref{eq:BIC}). One is the goodness of fit to the sample data, which is
determined by the negative model log-likelihood term. Another term is
the penalty of model complexity that is related to the number of model
parameters. The BIC value and its two contributing terms are listed in
Table \ref{tab:table2} for two models on image No. 13 .  The BIC value
of the K-method is 888.84, which is larger than that of the G-method. By
examining their individual contributing terms, we see that the
difference in their model complexity penalty terms are relatively minor
as compared to that of the model negative log-likelihood term. It shows
that the K-method does not offer a good model for JND points. 

\begin{table}[htbp]
\centering
\caption{Comparison of the model negative log-likelihood term, the model
complexity term and the BIC values of the K-method and the G-method for
Image No.  13.}\label{tab:table2}
\begin{tabular}{r|ccc}
\toprule
      & $-2ln(P(x|\Theta))$ & Complexity & BIC \\
\midrule
K-method & 824.65 & 64.19 & 888.84 \\
G-method & 485.37 & 91.70 & 577.07 \\
\bottomrule
\end{tabular}%
\end{table}%

We compare the performance of two methods for Image No. 13 side by side
in Fig. \ref{fig:jnd_example}. The first, second and third rows of the
figure display the input JND histograms before modeling, the output JND
histograms after modeling, and the final SQFs. The difference in
histograms in Fig.  \ref{fig:jnd_example} (a) and (b) is due to a
weighting scheme used in the K-method.  In that method, the JND point from
each individual was first normalized by the total observed JND numbers.
(For example, if a person observes $N$ JND points, each JND point of
him/her is weighted by $1/N$.) This scheme penalizes observed JND data
in the high QF region since people who observe JND points in the high QF
region tend to have a larger total JND number. In contrast, we do not
perform any weighting on the collected JND data. The JND histogram in
Fig. \ref{fig:jnd_example} (b) is obtained with raw user data. 

The output JND histograms obtained by the two methods are shown Fig.
\ref{fig:jnd_example} (c) and (d), respectively. In the K-method, the
location of the output JND point is determined as the median of JND
points within that cluster. These points are marked by circles in Fig.
\ref{fig:jnd_example} (a). This may lead to inaccurate result when JND
points are not clustered correctly. For example, the first four JND
points (from the left) in Fig. \ref{fig:jnd_example} (c) are very close
to each other while the last JND points is very far apart from the
others.  In our proposed method, the output JND locations are set to the
means of all Gaussian components. They are more stable and set apart
with more uniform spacing. 

It is worthwhile to point out that the total JND number for a given
image in the K-method is a pre-set number. This ad hoc choice has a
negative impact on its output JND histogram as shown in Fig.
\ref{fig:jnd_example} (c). It has 5 peaks only, which is less than the
G-method by two. In the proposed G-method, the JND number is determined
by optimizing the model with the lowest BIC value. Finally, Fig.
\ref{fig:jnd_example} (e) and (f) show the K-SQF and the G-SQF,
respectively. The G-SQF offers more quality levels, which is actually
more reasonable by re-examining the full set of JPEG coded images for
Image No.  13. 

\section{Conclusion and Future Work}\label{sec:copyright}

In this work, we proposed the use of the GMM to model the raw JND data
collected by the subjective test, and derived the G-SQF.  The new method
always provides a model that has a smaller BIC value than that proposed
in \cite{MCLJND}, indicating that it is a better model. The G-SQF for
all 50 source images will be made available to the public soon. It will
provide a training dataset of human perceptual experience on JPEG
images. We will develop a machine learning technique to predict viewer
experience on JPEG images that are not in the training dataset. 

\bibliographystyle{IEEEbib}
\bibliography{jnd}
\end{document}